\documentclass[conference]{IEEEtran}
\IEEEoverridecommandlockouts
\usepackage{cite}
\usepackage{amsmath,amssymb,amsfonts}
\usepackage{algorithmic}
\usepackage{graphicx}
\usepackage{textcomp}
\usepackage{xcolor}
\usepackage[acronyms]{glossaries}
\usepackage{listings}
\usepackage{booktabs}
\usepackage{array}
\usepackage{multirow}
\usepackage{soul,color}

\def\BibTeX{{\rm B\kern-.05em{\sc i\kern-.025em b}\kern-.08em
    T\kern-.1667em\lower.7ex\hbox{E}\kern-.125emX}}

\begin{document}

\title{Experimental Evaluation of Joint Clock Recovery and Equalization for Sub-Terahertz Links}

\author{\IEEEauthorblockN{
Pietro Savazzi\IEEEauthorrefmark{1},
Anna Vizziello\IEEEauthorrefmark{1},
Sherif Badran\IEEEauthorrefmark{2}, and
Josep M. Jornet\IEEEauthorrefmark{2}
}
\IEEEauthorblockA{\IEEEauthorrefmark{1}Department of Electrical, Computer and Biomedical Engineering, University of Pavia, Pavia, Italy\\
}
\IEEEauthorblockA{\IEEEauthorrefmark{2}Department of Electrical and Computer Engineering, Northeastern University, Boston, MA, USA\\
E-mail: \{pietro.savazzi, anna.vizziello\}@unipv.it\IEEEauthorrefmark{1}, \{badran.s, j.jornet\}@northeastern.edu\IEEEauthorrefmark{2}}
}

\newacronym{CFO}{CFO}{carrier frequency offset}
\newacronym{CR}{CR}{clock recovery}
\newacronym{BB}{BB}{baseband}
\newacronym{QAM}{QAM}{quadrature amplitude modulation}
\newacronym{IF}{IF}{intermediate frequency}
\newacronym{IQ}{IQ}{in-phase and quadrature}
\newacronym{LO}{LO}{local oscillator}
\newacronym{PN}{PN}{phase noise}
\newacronym{TED}{TED}{timing error detector}
\newacronym{CMA}{CMA}{constant modulus algorithm}
\newacronym{DPLL}{DPLL}{digital phase-locked loop}
\newacronym{DD}{DD}{decision directed}
\newacronym{NCO}{NCO}{numerically controlled oscillator}
\newacronym{ISI}{ISI}{intersymbol interference}
\newacronym{RF}{RF}{radio frequency}
\newacronym{BER}{BER}{bit error rate}
\newacronym{EVM}{EVM}{error vector magnitude}
\newacronym{MIMO}{MIMO}{multiple-input multiple-output}
\newacronym{PSG}{PSG}{performance signal generator}
\newacronym{AWG}{AWG}{arbitrary waveform generator}
\newacronym{DSO}{DSO}{digital storage oscilloscope}
\newacronym{PA}{PA}{power amplifier}
\newacronym{LNA}{LNA}{low-noise amplifier}
\newacronym{AMC}{AMC}{active multiplier chain}
\newacronym{DSP}{DSP}{digital signal processing}
\newacronym{ML}{ML}{maximum likelihood}
\newacronym{AWGN}{AWGN}{additive white Gaussian noise}
\newacronym{PLL}{PLL}{phase-locked loop}
\newacronym{OEPLL}{OEPLL}{optoelectronic phase-locked loop}
\newacronym{THz}{THz}{terahertz}
\newacronym{RRC}{RRC}{root raised cosine}

\maketitle

\begin{abstract}
This paper proposes and experimentally evaluates a joint clock recovery (CR) and equalization architecture tailored for high-speed sub-terahertz (sub-THz) wireless communication links. Specifically, a Baud-spaced digital receiver architecture is investigated that combines a constant modulus algorithm (CMA) equalizer with a blind timing error detector (TED), enabling robust symbol timing synchronization without decision-directed (DD) feedback or pilot symbols. The proposed TED leverages the CMA filter coefficients to estimate timing errors, which are then used to drive a Farrow interpolator operating at twice the symbol rate. The system is validated experimentally using a 140~GHz wireless testbed with 16-QAM modulation over a 10~GHz bandwidth. Results show that the proposed TED schemes outperform conventional blind TEDs, such as Gardner and blind implementations of Mueller \& M\"uller,  in terms of bit error rate (BER), error vector magnitude (EVM), and intersymbol interference (ISI) suppression. These capabilities are especially relevant to next-generation spaceborne communication systems, where wideband sub-THz links are expected to play a key role in enabling ultra-high-data-rate inter-satellite and deep-space communications under challenging synchronization constraints.
\end{abstract}


\glsresetall

\section{Introduction}
\label{sec:Introduction}

Sub-\gls{THz} communication systems operating in the 0.1--1~THz frequency range are emerging as a critical enabling technology for next-generation wireless networks, offering unprecedented data rates and bandwidth capabilities~\cite{Jiang2024}. This frequency regime is particularly attractive for spaceborne communication systems, where ultra-high-data-rate inter-satellite links and deep-space communications are essential for future space missions requiring massive data throughput, such as Earth observation constellations, lunar communications, and Mars exploration programs~\cite{Kokkoniemi2021,Tekbiyik2022}.

However, the implementation of sub-\gls{THz} communication systems faces significant technical challenges, particularly in the areas of synchronization and signal processing. At these frequencies, traditional synchronization approaches encounter several limitations. First, the increased \gls{PN} characteristics of sub-\gls{THz} oscillators severely impact system performance~\cite{Parisi2024}, as \gls{PN} effects scale quadratically with carrier frequency~\cite{Deng2022b}. Second, the wide bandwidths employed in sub-\gls{THz} systems---often spanning several gigahertz---exacerbate timing synchronization requirements and \gls{ISI} effects. Third, spaceborne applications impose additional constraints, including power limitations, radiation tolerance requirements, and the need for autonomous operation without ground-based reference signals~\cite{Kokkoniemi2021,Torrens2024}.
Moreover, for high-frequency space communications, accurate time and frequency synchronization play a very crucial role due to long propagation delays, high relative velocities, and the use of ultra-broadband signals. On the one hand, precise timing alignment ensures correct frame and symbol detection, enabling reliable demodulation and minimizing bit error rates. On the other hand, frequency synchronization compensates for Doppler-induced carrier frequency offsets and local oscillator drift, thereby maintaining phase coherence, which is essential for sophisticated modulation schemes and phase-sensitive signal processing techniques, such as beamforming. Thus, time and frequency synchronization are critical for robust and efficient operation of satellite and deep-space links, and reliable end-to-end communication under extreme space conditions.

\Gls{CR} and channel equalization represent two of the most critical signal processing functions in any high-speed communication receiver. Conventional approaches typically implement these functions sequentially, performing \gls{CR} first, followed by adaptive equalization. However, this separation can lead to suboptimal performance, particularly in challenging propagation environments where timing errors and channel distortions are strongly coupled. Moreover, traditional \glspl{TED} often rely on decision-directed feedback or pilot symbols~\cite{Jablon1992,Guo2023}, introducing latency and complexity that may be prohibitive in space applications where hardware resources are constrained.

Recent works in timing recovery have explored blind \gls{TED} architectures~\cite{Tang2024}, as well as classic schemes such as Gardner~\cite{Gardner1986} and Mueller \& M\"uller~\cite{Mueller1976}. A modern, hardware‑efficient Baud‑rate \gls{TED} was proposed~\cite{Savazzi2008}, further motivating low‑complexity timing recovery without pilot‑based feedback. Early studies on parallel timing recovery with adaptive equalization~\cite{Xian2011} also support the motivation for joint \gls{CR}/equalization approaches.

This paper addresses these challenges by introducing and experimentally validating a novel joint clock recovery and equalization architecture~\cite{Jablon1992} specifically designed for sub-\gls{THz} wireless communication links. The key innovation lies in the tight integration of \gls{CMA} equalization with blind \gls{TED}~\cite{Gardner1986}, where the equalizer impulse response coefficients are directly leveraged to estimate timing errors. Unlike conventional approaches that separate \gls{CR} and equalization functions, the proposed method exploits the inherent relationship between equalizer tap weights and timing-induced \gls{ISI} to achieve superior synchronization performance.
The proposed timing error detector extracts timing information directly from the \gls{CMA} filter coefficients without requiring symbol decisions or pilot sequences, thereby reducing computational complexity and eliminating the need for a demapper in the timing recovery loop. The timing error estimates drive a Farrow interpolator~\cite{Farrow1988} operating at twice the symbol rate, enabling precise fractional delay adjustments. In its simplified implementation, the algorithm requires only a single addition operation per symbol, representing a significant complexity reduction compared to conventional \glspl{TED} that requires multiplications. Experimental validation is performed using the TeraNova sub‐\gls{THz} testbed~\cite{Priyangshu2020} at 140~GHz with 16-QAM over a 10~GHz bandwidth. Other similar experimental works~\cite{Shadi2021,Wang2013} proved the hardware feasibility of these links.

The main contributions of this work are threefold: (1) the development of a novel joint \gls{CR}/\gls{CMA} architecture that achieves superior timing synchronization performance compared to traditional blind \glspl{TED}; (2) experimental validation using a 140~GHz wireless testbed with 16-QAM modulation over a 10~GHz \gls{IF} bandwidth, demonstrating the practical feasibility of the approach; and (3) comprehensive performance analysis showing improvements in \gls{BER}, \gls{EVM}, and \gls{ISI} suppression compared to state-of-the-art blind timing recovery algorithms, including Gardner and Mueller \& M\"uller \glspl{TED}.

The experimental results confirm that the proposed scheme outperforms conventional blind \glspl{TED} across all evaluated metrics while requiring minimal computational resources. These capabilities make the approach particularly suitable for next-generation spaceborne communication systems, where the combination of wide bandwidth, stringent power constraints, and autonomous operation requirements demands both high-performance and low-complexity synchronization solutions.

The remainder of this paper is organized as follows: Sec.~\ref{sec:SystemModel} describes the system model and signal formulation, Sec.~\ref{sec:CRCMAscheme} details the proposed joint \gls{CR}/\gls{CMA} scheme and timing error detector derivation, Sec.~\ref{sec:THzTestbed} presents the experimental sub-\gls{THz} testbed and implementation details, Sec.~\ref{sec:Results} provides comprehensive experimental results and performance comparisons, and finally, Sec.~\ref{sec:Conclusion} concludes the paper and outlines future research directions.

\section{System Model}
\label{sec:SystemModel}
In the considered system architecture, the receiver implements equalization, clock, and \gls{CFO} recovery at \gls{BB}. 
The \gls{IF} modulated signal can be represented as:
\begin{equation}
\label{eq:s_IF}
s_{t,IF}(t)=\Re\left\{s_t(t)e^{j2\pi f_{IF}t}\right\},
\end{equation}
with $f_{IF}=5$~GHz, and
\begin{equation}
	\label{baseband signal}
    s_t(t)=\sum_{n}c_ng(t-nT),
\end{equation}
where $c_n$ are the complex \gls{QAM} symbols, $g(t)$ is the pulse shaping filter, e.g., a \gls{RRC} filter, $T$ is the symbol interval. 
The \gls{IF} signal is then upconverted to \gls{RF} as described in:
\begin{equation}
    \label{eq:RFModulatedSignal}
    s_{t,RF}(t) = s_{t,IF}(t) \cos\big( 2\pi f_{ct} t+ \phi_t(t) \big),
\end{equation}
where $f_{ct}=140$~GHz, and $\phi_t(t)$ is the random \gls{PN} of the transmit oscillator.

\begin{figure*}
    \centering
    \includegraphics[width=0.7\linewidth]{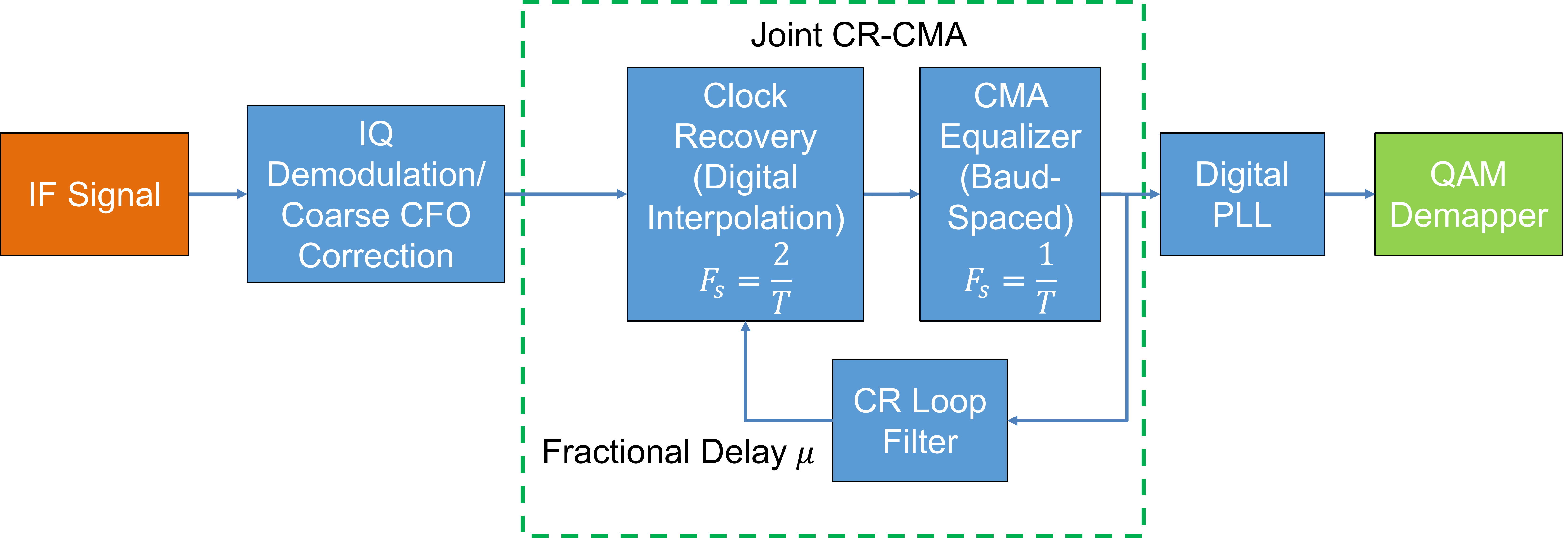}
    \caption{A block diagram depicting the receiver scheme.}
    \label{fig:RxScheme}
\end{figure*}

The received \gls{IF} signal can be modeled as:
\begin{equation}
    \label{eq:ReceivedIF}
    s_{r,IF}(t)=\Re\big\{s_{t,IF}(t)e^{j\Delta \theta(t)}\big\},
\end{equation}
with 
\begin{equation}
\label{eq:Phase_DifferenceBB}
\Delta \theta(t) = \theta_t(t) - \theta_r(t)=2\pi \Delta f_c t+\Delta \phi(t),
\end{equation}

\begin{equation}
\label{eq:Freq_DifferenceBB}
\Delta f_c = f_{ct} - f_{cr},
\end{equation}

\begin{equation}
\label{eq:Phase_Noise}
\Delta \phi(t) = \phi_t(t) - \phi_r(t),
\end{equation}
where $\Delta f_c$ is the resulting \gls{CFO}, while $\Delta \phi(t)$ represents the received \gls{PN}. 

The \gls{BB} signal is then computed as:
\begin{equation}
\label{eq:s_rBB}
    s_r(t)=s_{r,IF}(t)e^{j(2\pi f_{IF}t+\phi_{BB})}.
\end{equation}
The \gls{BB} signal in Eq.~\eqref{eq:s_rBB} can be written as:
\begin{equation}
    \label{eq:s_rBBvsPhaseError}
    s_r(t)=\big(\sum_{n}c_n \rho(t-nT-\tau(t))\big)e^{j\Delta \theta_r(t)},
\end{equation}
where $\tau(t)$ represents the timing error with respect to $T$ the symbol time duration, $\Delta \theta_r(t)$ corresponds to including in $\Delta \theta(t)$ the phase term $\phi_{BB}$ due to the non-perfect time synchronization between the transmitter and receiver, and $\rho(t)$ is the autocorrelation of $g(t)$, i.e., the \gls{RRC} filter used at both the transmit and receive sides. As explained in Sec.~\ref{sec:THzTestbed}, the transceiver signal processing from \gls{BB} to \gls{IF} is implemented in MATLAB, the IQ modulation and demodulation are performed numerically at a sample rate corresponding to 32 samples per symbol.

\gls{CFO} recovery and \gls{PN} filtering have been implemented by a \gls{DPLL} and will be deeply discussed in a future paper. In this work, we focus on the joint \gls{CR} and equalization scheme as described in the next section and depicted in Fig.~\ref{fig:RxScheme}.

\section{Proposed Joint CR/CMA Scheme}
\label{sec:CRCMAscheme}
The received \gls{IF} signal is sampled at $F_s=\frac{32}{T}=160$~GSa/s and \gls{IQ} demodulated, providing the signal $s_r(t)$.
After \gls{IQ} demodulation, the \gls{CR} is implemented by a digital Farrow interpolator~\cite{Farrow1988,Savazzi2008}, whose input fractional delay is the output of a first-order loop filter with a blind \gls{TED} that is obtained from the \gls{CMA} equalizer impulse response as explained in the following, and highlighted in the Joint \gls{CR}-\gls{CMA} section of Fig.~\ref{fig:RxScheme}.

The input of the \gls{CR} interpolation filter is $x(n)=s_r(n\frac{T}{2})$.

The output of the \gls{TED} at the time instant $t=nT$ is defined as $\epsilon(n)$, while the output of the first-order loop filter is computed as:
\begin{equation}
    \label{eq:CRLFoutput}
    \mu(n)=\mu(n-1)+\alpha_c \epsilon(n-1),
\end{equation}
where $\mu(n)$ represents a fractional delay that can be used as the main input to the digital interpolator~\cite{Savazzi2008}, while $\alpha_c$ is a suitable step size or learning factor. It functions as a \gls{NCO}, as shown in~\cite{Erup1993}, by managing both delays and advances, and $\epsilon(n)$ will be derived in the subsequent Eq.~\eqref{eq:ProposedTED}.

It is important to note that this digital implementation could be replaced by an analog-controlled oscillator, driven by the same input coming from Eq.~\eqref{eq:CRLFoutput}. This hybrid approach could be more feasible at the operating bandwidth. In this sense, the presented results can be thought of as a digital simulation of the mixed analog/digital implementation.

The samples at the output of the digital interpolator are sampled at the symbol time (Baud spaced):
\begin{equation}
\label{equ:BBRxSignalVector}
    \mathbf{r}(n)=\left[r(n),r(n-1),\dots,r(n-p+1)\right]^T,
\end{equation}
where $\mathbf{r}(n)$ denotes the vector of Baud-spaced samples, computed at time $nT$.

The received signal at the output of the \gls{CMA} equalizer is:
\begin{equation}
    \label{equ:CMAequOut}
    \mathbf{x}(n)=\mathbf{w}^T_n\mathbf{r}(n),
\end{equation}
with
\begin{equation}
\label{equ:EquCoeff}
    \mathbf{w}_n=\left[w_n(0),w_n(1),\dots,w_n(p-1)\right]^T,
\end{equation}
where $\mathbf{w}(n)$ denotes the equalizer filter coefficient vector at time $nT$, and $p$ corresponds to the filter length.

The instantaneous phase at the equalizer output is then corrected, according to the \gls{DPLL} output:
\begin{equation}
    \label{equ:EquOutPhaseCorrected}
    y(n)=x(n)e^{-j\hat{\theta}(n)},
\end{equation}
where 
 \begin{equation}
\label{equ:EstimatedPhaseDiscretTime}
    \hat{\theta}(n)\triangleq \hat{\Delta\theta}(nT),
\end{equation}
and $\hat{\Delta\theta}(t)$ is the estimate of the previously defined $\Delta\theta(t)$, that includes both \gls{CFO} and \gls{PN} effects.

The coefficients of the equalizer are then updated by using:
\begin{equation}
    \label{equ:CMAcoeffUpdate}
    \mathbf{w}_{n+1} = \mathbf{w}_n - \alpha_e \left( |y(n)|^2 - R \right) y(n) \mathbf{r}^*(n),
\end{equation}
with $\alpha_e$ a suitable step size, i.e., the equalizer learning factor, while:
\begin{equation}
    \label{R_CMA}
    R = \frac{E\{|c_n|^4\}}{E\{|c_n|^2\}}.
\end{equation}

\subsection{The Proposed TED}

Since both \gls{CMA} equalization and \gls{DPLL} Baud-spaced schemes require perfect timing to work properly, the \gls{CR} loop and \gls{TED} are implemented before them~\cite{Tang2024,Xu2024}, with a \gls{TED} that does not use symbol decisions at the output of the \gls{QAM} demapper~\cite{Jablon1992} or use symbol decisions and/or pilot symbols~\cite{Guo2023}. 

In this work, we use the equalizer impulse response to compute a suitable \gls{TED} function, effectively merging the \gls{CMA} loop with the \gls{CR} one.
The main idea comes from the fact that a \gls{TED} can be constructed by minimizing some coefficients of the equivalent \gls{CR} output impulse response that represents \gls{ISI} due to the non-perfect symbol timing~\cite{Mueller1976}.

Following this reasoning, the input of the \gls{CR} filter loop is derived from the \gls{CMA} filter coefficients $\mathbf{w}_{n}$ as follows:
\begin{equation}
    \label{eq:ProposedTED}
    \epsilon(n)=-\sum_{k\neq i}\Re\{w_k(n)\},
\end{equation}
where $\Re$ denotes the real part of its argument, and $i=\frac{p+1}{2}$ represents the central coefficient of the \gls{CMA} filter, with odd $p$.

The meaning of Eq.~\eqref{eq:ProposedTED} comes from the fact that the \gls{CR} scheme must reduce the \gls{ISI} related to non-perfect symbol sampling. In this sense, using Eq.~\eqref{eq:ProposedTED} as the input of the \gls{CR} loop filter aims to minimize the \gls{ISI} due to non-perfect symbol timing estimation, allowing the equalizer to counteract the \gls{ISI} coming from other channel or \gls{IF} and \gls{RF} impairments.

Eq.~\eqref{eq:ProposedTED} could also be implemented as:
\begin{equation}
    \label{eq:AlternativeTED}
    \epsilon(n)=-\left(\sum_{k\neq i}\Re\{w_k(n)\}+\sum_{k\neq i}\Im\{w_k(n)\}\right),
\end{equation}
with $\Im$ denoting the imaginary part of the argument. Actually, the use of the complex Eq.~\eqref{eq:AlternativeTED} does not provide any advantage over the use of Eq.~\eqref{eq:ProposedTED}, since \gls{DPLL} phase correction makes the real component of $w_n(i)$ predominant over its imaginary part.

Moreover, a simplified implementation of Eq.~\eqref{eq:ProposedTED} can be derived by only considering the two coefficients adjacent to the central one:
\begin{equation}
    \label{eq:modifiedCMAbasedTED}
    \epsilon(n)=-(\Re\{w_k(i-1)\}+\Re\{w_k(i+1)\}),
\end{equation}
where $i=\frac{p+1}{2}$. This simplified implementation is reasonable since most of the energy is in the central equalizer coefficients.

The proposed \gls{TED} performances will be compared with the commonly used ones, considering the algorithms summarized in~\cite{Tang2024} that do not make use of the demapper symbol decisions, by also measuring the energy reduction of the steady-state equalizer coefficients, that is related to the ability of the \gls{CR} algorithm to reduce the \gls{ISI} level due to non-perfect timing.
Other \gls{DD} algorithms were not taken into account in order to reduce latency estimation, especially if considering the \gls{CR} hybrid analog/digital implementation discussed previously.

\section{Terahertz Testbed and Receiver Implementation}
\label{sec:THzTestbed}

\begin{figure*}
    \centering
    \includegraphics[width=0.7\linewidth]{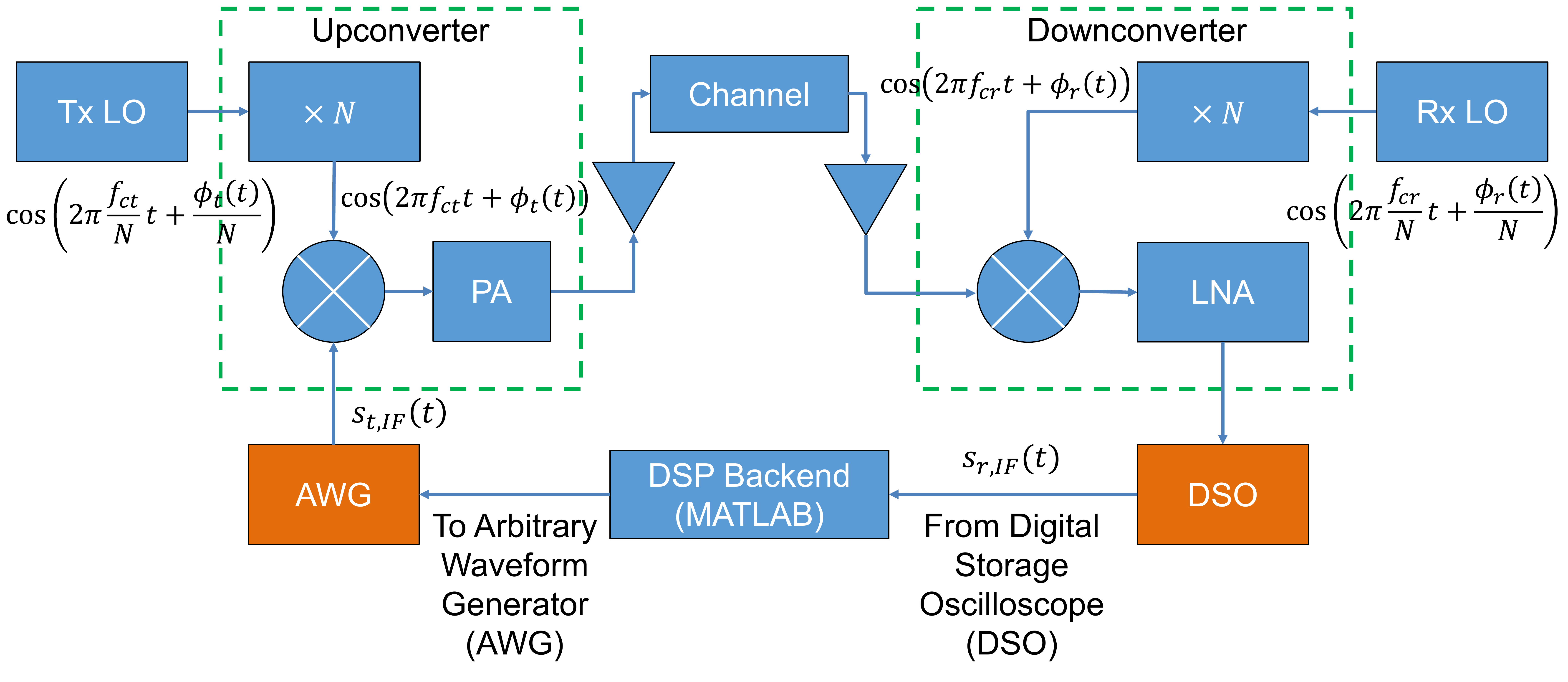}
    \caption{A block diagram depicting the full sub-terahertz transceiver testbed.}
    \label{fig:TestBedScheme}
\end{figure*}

\begin{figure}
    \centering
    \includegraphics[width=\linewidth]{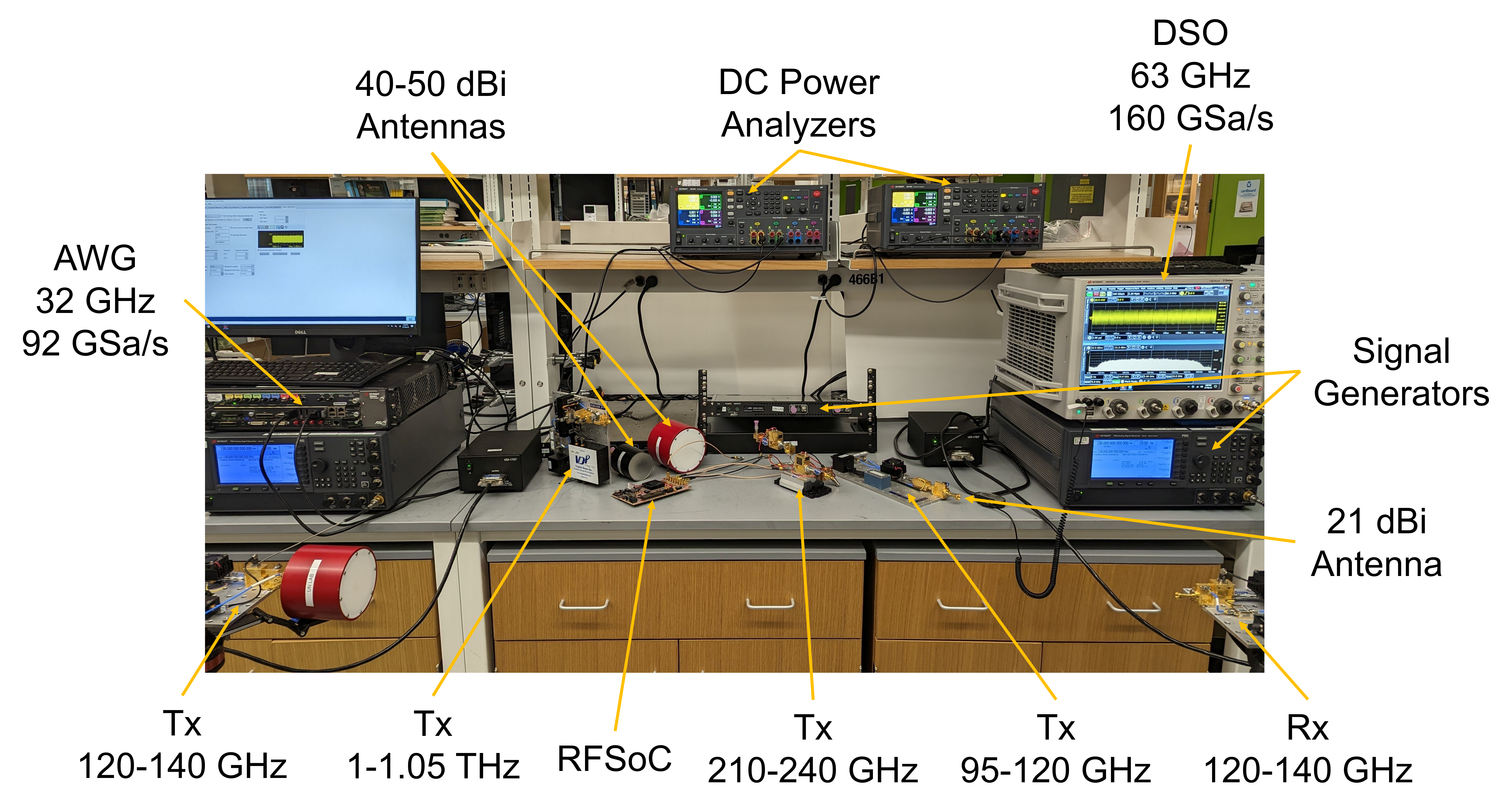}
    \caption{A picture of the actual testbed.}
    \label{fig:testbed picture}
\end{figure}

For experimental validation, we utilized our TeraNova terahertz testbed~\cite{Priyangshu2020}, which is a very versatile and modular testbed as depicted in Fig.~\ref{fig:TestBedScheme}. It is designed to operate in various frequency bands ranging from 95~GHz to 1.05~THz, and due to its modularity, operation in different frequency ranges can be realized by simply swapping the up and downconverters and antennas with different ones, as depicted in Fig.~\ref{fig:testbed picture}. The measurements in this paper were performed at a center \gls{RF} frequency of 140~GHz.

The transmitter comprises a D-band (110-170~GHz) upconverter and high-gain (21~dBi) horn antenna from Virginia Diodes, Inc., and a \gls{PSG} (50~GHz) and an \gls{AWG} (32~GHz, 92~GSa/s) from Keysight Technologies. This upconverter consists of an \gls{AMC} that multiplies the frequency of the input \gls{LO} signal (from the \gls{PSG}) by a factor of 4, a mixer, and a high-gain (20~dB) \gls{PA}. The transmitter takes an \gls{IF} signal (data-bearing modulated signal) and upconverts it into an \gls{RF} signal in the sub-terahertz or terahertz frequency range.

The receiver comprises a D-band (110-170~GHz) downconverter and a similar horn antenna from Virginia Diodes, Inc., and a similar \gls{PSG} and a \gls{DSO} (63~GHz, 160~GSa/s) from Keysight Technologies. This downconverter consists of a similar mixer and \gls{AMC} with the same multiplication factor of 4, and a high-gain (11~dB) \gls{LNA}. The receiver takes an \gls{RF} signal in the sub-terahertz or terahertz frequency range and downconverts it into an \gls{IF} signal that can be digitized and post-processed.

We have our entire \gls{DSP} backend developed in MATLAB. We export our designed \gls{IF} discrete-time modulated information-bearing signals from MATLAB and upload them onto the \gls{AWG}, which converts them into continuous-time analog \gls{IF} signals that feed the \gls{IF} input port of the upconverter. Similarly, the \gls{IF} output port of the downconverter feeds the \gls{DSO}, which digitizes (i.e., samples, quantizes, and stores) the received analog \gls{IF} signal. We then import the received digital \gls{IF} signal back to MATLAB for further signal processing.

\section{Experimental Results}
\label{sec:Results}
Frames of 21,010 16-QAM symbols were acquired, i.e., $s_{r,IF}(t)$ at the output of the receiver \gls{IF} port as represented in Fig.~\ref{fig:TestBedScheme}. The main testbed system parameters are summarized in Table~\ref{tab:TestbedSysPar}, while the main \gls{BB} parameters are shown in Table~\ref{tab:BBSysPar}. The \gls{CMA} equalizer converges after the first received 8,000 bits, corresponding to 2,000 QAM symbols. The presented results thus refer to around 19,000 16-QAM received symbols.

\begin{table}
\caption{Testbed System Parameters }\label{tab:TestbedSysPar}
\centering
\begin{tabular*}{\columnwidth}{@{\extracolsep{\fill}}lr}
\toprule
Modulation scheme & 16-QAM\\

\gls{RF} carrier frequency & 140 GHz\\

\gls{IF} carrier frequency & 5 GHz\\

IF bandwidth & 10 GHz\\

$F_s$ (\gls{IF} sampling frequency) & 160 GSa/s\\

$f_s$ (\gls{BB} sampling frequency) & 10 GSa/s\\
\bottomrule
\end{tabular*}
\end{table}

\begin{table}
\caption{Baseband System Parameters}\label{tab:BBSysPar}
\centering
\begin{tabular*}{\columnwidth}{@{\extracolsep{\fill}}lr}
\toprule
CMA filter length $p$ & 21\\

CMA step size $\alpha_e$ & $9\cdot 10^{-4}$\\

TED step size $\alpha_c$ for independent TEDs & $1\cdot 10^{-2}$\\

CMA-based TEDs step size $\alpha_c$ & $1.3\cdot 10^{-4}$\\
\bottomrule
\end{tabular*}
\end{table}

\begin{figure}
    \centering
    \includegraphics[width=0.7\linewidth]{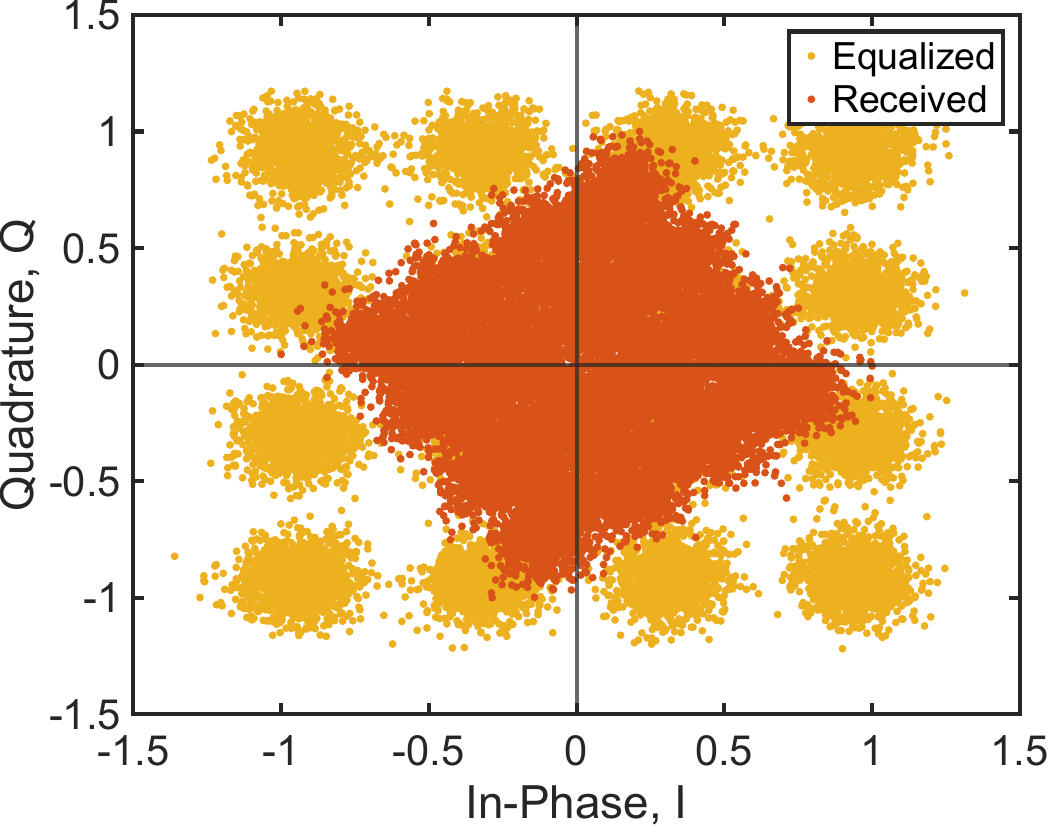}
    \caption{The received constellation diagram before and after equalization.}
    \label{fig:RxScatterPlot}
\end{figure}

In Fig.~\ref{fig:RxScatterPlot}, the constellation diagram of the equalized received signal, obtained with the proposed \gls{TED} in Eq.~\eqref{eq:ProposedTED}, is shown.

In Table~\ref{tab:TedPerformanceComparison}, the proposed \gls{CMA}-based \glspl{TED} of Eq.~\eqref{eq:ProposedTED} and Eq.~\eqref{eq:modifiedCMAbasedTED} are compared with the ones described in~\cite{Tang2024,Gardner1986}, in terms of \gls{BER}, \gls{EVM}, and the normalized energy of the \gls{CMA} impulse response due to the residual \gls{ISI} after \gls{CR} and caused by other channel impairments.
In more detail, the normalized \gls{ISI} energy is defined as follows:
\begin{equation}
    \label{eq:EnergyReduction}
    \frac{\sum_{k\neq i}|w_k(L)|^2}{\sum_k |w_k(L)|^2},
\end{equation}
with $i=\frac{p+1}{2}$, and $L$ being the total simulation length.

\begin{table}
\caption{TED Performance Comparison}\label{tab:TedPerformanceComparison}
\centering
{\fontsize{7.5}{10}\selectfont
\begin{tabular*}{\columnwidth}{@{\extracolsep{\fill}}cccc}
\toprule
\textbf{TED Algorithm} & \textbf{BER} & \textbf{EVM} & \textbf{Residual ISI}\\
\midrule
CMA-based, Eq.~\eqref{eq:ProposedTED} & $3.82\cdot 10^{-4}$ & 34.01\% & 0.0677 \\

Modified CMA-based, Eq.~\eqref{eq:modifiedCMAbasedTED} &  $3.82\cdot 10^{-4}$ & 33.98\% & 0.0678\\

Gardner~\cite{Gardner1986} & $4.21\cdot 10^{-4}$ & 34.24\% & 0.0937\\ 

ABS TED~\cite{Tang2024} & $5.53\cdot 10^{-4}$ & 34.04\% & 0.759\\

Sign-MM TED~\cite{Tang2024} & $6.32\cdot 10^{-4}$ & 34.09\% & 0.0691 \\

Modified ABS-TED~\cite{Tang2024} & $6.9\cdot 10^{-2}$ & 47.71\% & 0.0861\\
\bottomrule
\end{tabular*}}
\end{table}

\begin{figure}
    \centering
    \includegraphics[width=0.7\linewidth]{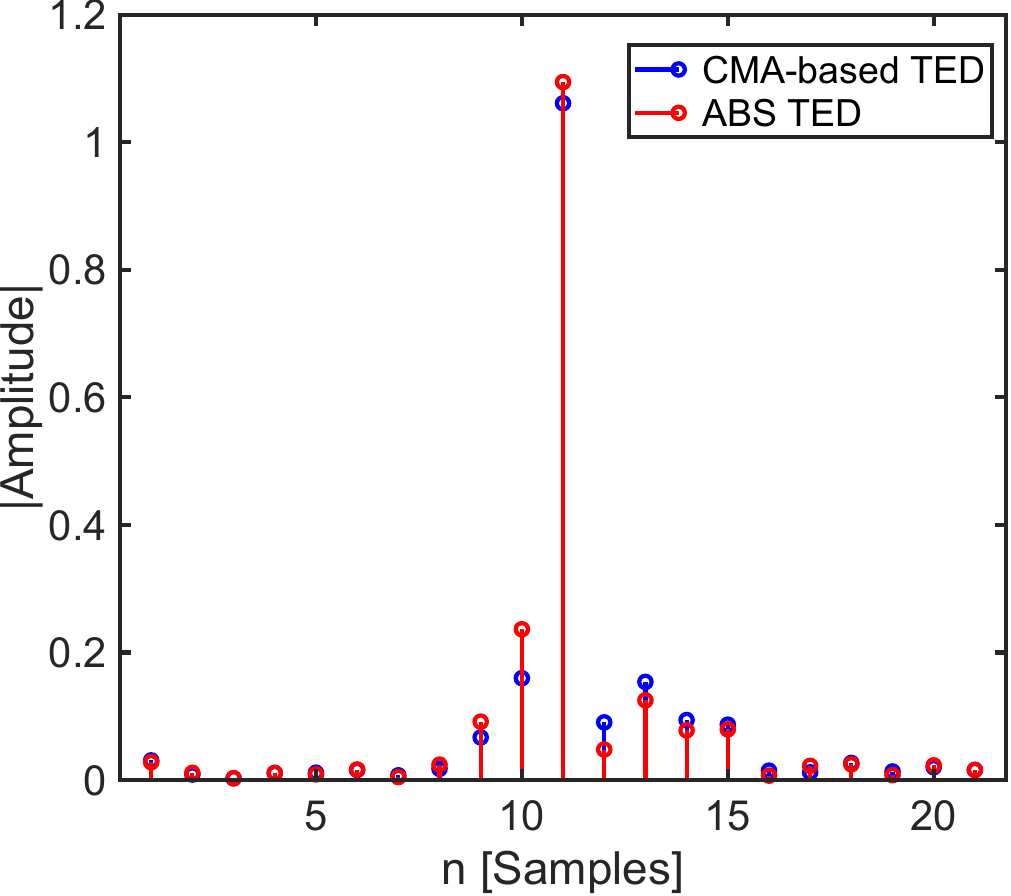}
    \caption{The modulus of the CMA filter coefficients with CMA-base and ABS TEDs.}
    \label{fig:EquCoeff}
\end{figure}

Both proposed \gls{TED} computation techniques outperform the other disjoint algorithms across all three metrics under study, providing less \gls{BER}, \gls{EVM}, and \gls{ISI} energy of the \gls{CMA} impulse response. It is important to highlight that the metric in Eq.~\eqref{eq:EnergyReduction} is reduced as long as the \gls{CR} scheme is able to perform symbol timing synchronization, reducing the level of \gls{ISI} to be minimized by the \gls{CMA} equalizer. This can also be appreciated by comparing the \gls{CMA} channel impulse response using the \gls{CMA}-based \gls{TED} and the ABS ones, as illustrated in Fig.~\ref{fig:EquCoeff}. The energy of the impulse response coefficients obtained with the \gls{CMA}-based \gls{TED} is lower than the ones corresponding to the ABS \gls{TED}.

The complexity burden of the proposed schemes, especially the one in Eq.~\eqref{eq:modifiedCMAbasedTED}, is comparable with the other schemes taken into account in Table~\ref{tab:TedPerformanceComparison}, except for the sign-MM \gls{TED}, i.e., the signed Mueller \& M\"uller \gls{TED}, that requires fewer bits to be implemented as it uses only the sign of the input samples~\cite{Tang2024}.

The proposed \gls{TED} schemes require a lower step size value than the one used for the other algorithms, since the \gls{CMA} impulse response includes the energy needed to counteract \gls{ISI} that does not come from the non-perfect symbol timing acquisition. Moreover, the convergence speed is comparable with the one of the considered blind \gls{TED} algorithms.

\subsection{Computational Complexity Comparison}

In Table~\ref{tab:ComplexityComparison}, a comparison of computational complexity of the algorithms is shown, in terms of the number of real multiplications and sums. It is important to highlight that in both the sign-MM and modified ABS \glspl{TED}, the needed multiplications actually correspond to changing the sign. However, the proposed scheme, especially if considering the simplest implementation that uses only two coefficients of the \gls{CMA} impulse response, has lower complexity. Moreover, the performance of the simpler modified \gls{CMA}-based is comparable with using $p-2$ multiplications as reported in Table~\ref{tab:TedPerformanceComparison}.

\begin{table}
\caption{Comparison of TED Algorithms by Number of Real Multiplications and Sums}
\label{tab:ComplexityComparison}
\centering
\begin{tabular*}{\columnwidth}{@{\extracolsep{\fill}}ccc}
\toprule
\textbf{TED Algorithm} & \textbf{Multiplications} & \textbf{Sums/Subtractions} \\
\midrule
Gardner & 2 & 3 \\

ABS TED & 1 & 2 \\

Sign-MM TED & 2 & 1 \\

Modified ABS TED & 2 & 1 \\

CMA-based & 0 & $p-2$ \\

Modified CMA-based & 0 & 1 \\
\bottomrule
\end{tabular*}
\end{table}

\section{Conclusion}
\label{sec:Conclusion}

This work has demonstrated the effectiveness of a joint clock recovery and \gls{CMA} equalization architecture for sub-\gls{THz} wireless communication systems, with particular attention to spaceborne applications. By exploiting the structure of the \gls{CMA} equalizer impulse response, the proposed blind \gls{TED} achieves robust timing synchronization without relying on pilot sequences or demapper decisions, which is an important advantage for reducing complexity and latency in space environments where hardware constraints, radiation tolerance, and power efficiency are critical. Experimental validation at 140~GHz with 16-QAM modulation confirms that the proposed architecture delivers improved \gls{BER}, \gls{EVM}, and \gls{ISI} mitigation compared to traditional blind \gls{TED} schemes. These results highlight the suitability of the proposed approach for future space communication systems operating in the sub-\gls{THz} band, such as inter-satellite or satellite-to-ground links, where precise timing synchronization over wideband channels is essential. Future research will focus on the analysis of the theoretical explanation of the proposed algorithm, incorporating carrier recovery and developing hybrid analog/digital implementations optimized for the stringent requirements of space-qualified platforms. Moreover, we will investigate the feasibility of applying such synchronization schemes to distributed \gls{MIMO} satellite communications~\cite{Savazzi2015,Zhang2025}, as precise synchronization is essential for achieving beam focusing in distributed \gls{MIMO} systems.

\section*{Acknowledgment}

This work was supported by the U.S. National Science Foundation (NSF) under Awards CNS-2225590 and CNS-2346487, and by the European Union under the Italian National Recovery and Resilience Plan (NRRP) of NextGenerationEU, partnership on “Telecommunications of the Future” (PE00000001 - program “RESTART”).

\bibliographystyle{IEEEtran}
\bibliography{references.bib}

\end{document}